\documentclass[preprint,12pt]{elsarticle}
\usepackage{amsmath,amssymb,graphicx,bm}
\usepackage{url}   %
\usepackage{tikz}
\usetikzlibrary{arrows.meta,positioning,fit,backgrounds,calc,shapes.geometric}

\newcommand{\Dh}{\ensuremath{D_\mathrm{H}}}
\tikzset{
  stage/.style={draw=blue!55!black, line width=0.7pt, rounded corners=2.5pt,
                align=center, inner sep=5pt, font=\small, fill=blue!7},
  gnnbox/.style={draw=orange!80!black, line width=1.0pt, rounded corners=2.5pt,
                 align=center, inner sep=5pt, font=\small, fill=orange!15},
  datab/.style={draw=green!45!black, line width=0.7pt, rounded corners=2.5pt,
                align=center, inner sep=4pt, font=\footnotesize, fill=green!9},
  flow/.style={-{Stealth[length=2.8mm,width=2.2mm]}, line width=0.9pt, draw=black!70},
  flab/.style={font=\scriptsize, text=black!70, fill=white, inner sep=1.5pt},
  numbadge/.style={circle, fill=blue!55!black, text=white,
                   font=\bfseries\footnotesize, inner sep=1.0pt, minimum size=4.6mm},
}

\graphicspath{{figs/}{./}}
\journal{Journal of Nuclear Materials}

\begin{document}
\begin{frontmatter}

\title{Graph neural network prediction of temperature-dependent hydrogen
diffusion and thermal conductivity tensors of tungsten containing helium
bubbles and grain boundaries}

\author[yu]{S.~Saito\corref{cor1}}
\ead{saitos@yz.yamagata-u.ac.jp}
\author[nifs]{M.I.~Kobayashi}
\author[yu]{T.~Kasahara}
\cortext[cor1]{Corresponding author.}
\affiliation[yu]{organization={Graduate School of Science and Engineering,
Yamagata University}, city={Yonezawa}, country={Japan}}
\affiliation[nifs]{organization={National Institute for Fusion Science},
city={Toki}, country={Japan}}

\begin{abstract}
Helium bubbles and grain boundaries produced in tungsten plasma-facing
components alter hydrogen-isotope transport and thermal conduction by orders
of magnitude, yet evaluating these transport properties for a given
microstructure requires hours of molecular dynamics (MD) per configuration.
We present a graph neural network (GNN) surrogate that maps an atomic
configuration of tungsten containing helium bubbles and grain boundaries
directly to the full $3\times3$ symmetric tensors of the hydrogen diffusion
coefficient $\Dh(T)$ and the thermal conductivity $\kappa(T)$ at arbitrary
temperature. Anisotropy is captured by a rotation-equivariant geometric
tensor pooling layer, and temperature is handled by predicting
temperature-independent physical parameters (an Arrhenius pair
$(D_0,E_a)$, a phonon conductivity tensor, and a defect residual
resistivity) that are expanded analytically through the Arrhenius and
Wiedemann--Franz--Matthiessen relations. Training labels for 635
microstructures (512 structures spanning four base classes plus a
trapping-focused extension)
are generated with an inexpensive embedded-atom-method
potential (Green--Kubo phonon conductivity and multi-temperature tracer
diffusion), the electronic conduction channel is calibrated against
published irradiation-degradation measurements, and the surrogate pipeline
is anchored to a first-principles machine-learning potential
(VASP+FLARE) through paired MD calibration runs and an active-learning
loop. The learned activation energies (median $0.21$\,eV, rising in
bubble and grain-boundary structures) reproduce literature hydrogen
migration barriers and trapping physics; the rotation-equivariant tensor
pooling keeps the model orientation consistent on arbitrarily oriented
sub-blocks (activation-energy coefficient of variation $0.2$--$0.3\%$), the
property that matters when a large specimen is queried block by block.
Coupled finite-element thermal--hydrogen analyses driven by the surrogate
show the conductivity degradation changing predicted hydrogen permeation by
a factor of $2.5$ through the temperature field. The model returns both
tensors in milliseconds, enabling microstructure-resolved transport input
for component-scale analyses of fusion divertor components.
\end{abstract}

\end{frontmatter}

\section{Introduction}
\label{sec:intro}

Tungsten is the leading plasma-facing material (PFM) for ITER and DEMO
divertors owing to its high melting point, high thermal conductivity and
low sputtering yield. Under operation, however, its near-surface
microstructure departs rapidly from the pristine crystal: helium from the
plasma and from $(n,\alpha)$ transmutation precipitates into nanometric
bubbles, displacement damage accumulates, and recrystallisation and
manufacturing routes leave dense grain-boundary (GB) networks. These
features control the two transport properties that matter most for a
divertor plate (target): the transport of hydrogen isotopes (hence tritium
retention and permeation) and heat conduction to the coolant. Experiments
show that thermal conductivity of irradiated or helium-implanted tungsten
degrades to roughly half of its pristine value already at
$\sim$0.5\,dpa or a few thousand appm of helium
\cite{Habainy2018,Sina2024,Reza2020}, while helium bubbles act as
strong traps that reduce effective hydrogen diffusivity by orders of
magnitude \cite{Wang2024effD}, and dense grain-boundary networks
contribute additional trapping. Component-scale thermo-mechanical and
tritium-transport analyses therefore need transport coefficients that are
\emph{local functions of the evolving microstructure}, not handbook values.

\begin{figure}[t]
\centering
\includegraphics[width=\linewidth]{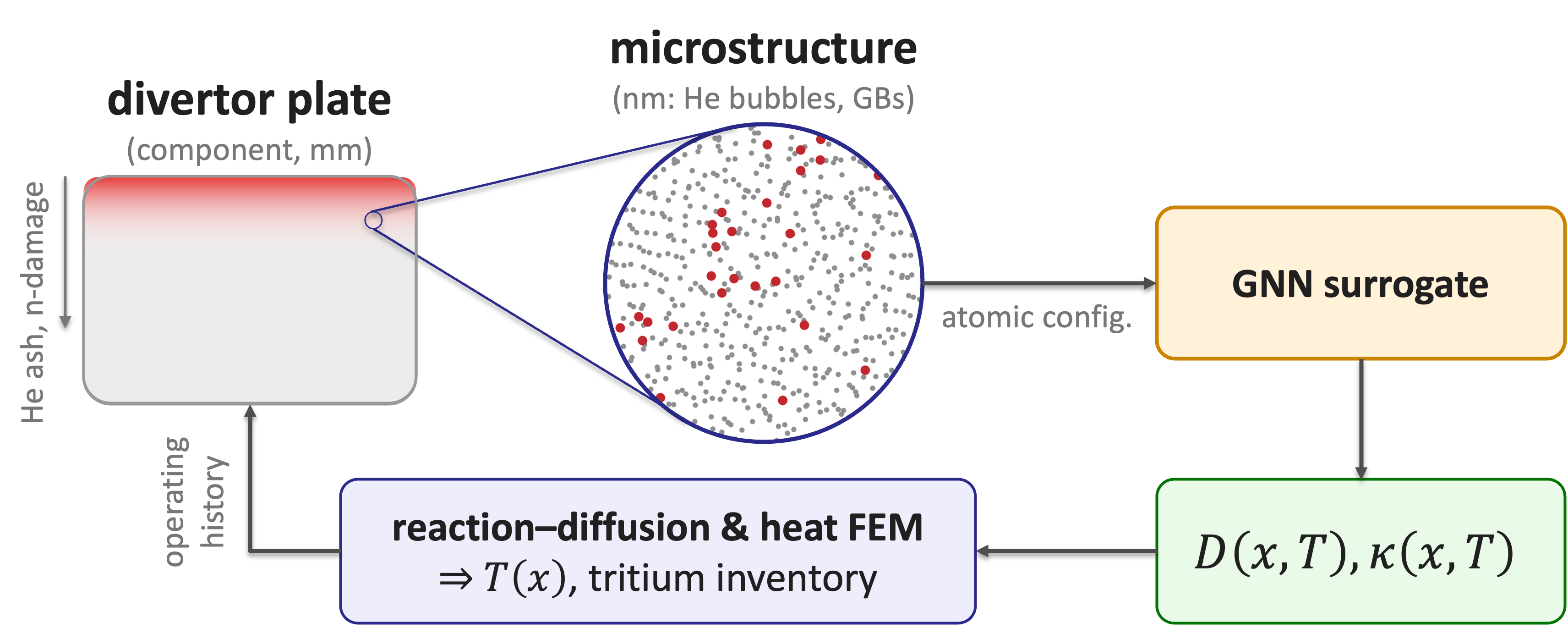}
\caption{Motivation. Component-scale prediction of tritium inventory and
temperature in a tungsten wall solves reaction--diffusion and heat
equations whose coefficients depend on the local, irradiation-evolved
microstructure. This work provides the missing link: a surrogate that
returns the local anisotropic diffusion and conductivity tensors
$\bm{D}(\mathbf{x},T)$, $\bm{\kappa}(\mathbf{x},T)$ directly from the atomic
configuration, closing the loop back into the continuum solver.}
\label{fig:motivation}
\end{figure}

Quantifying the tritium inventory and the heat handling of such a wall is,
in practice, a boundary-value problem: the fuel and heat balances are
obtained by solving reaction--diffusion and heat-conduction equations over
the component, driven by source terms from the implanted helium ash and the
$(n,\alpha)$/displacement damage that accumulate under 14\,MeV neutron
irradiation \cite{Kobayashi2013,Hino2013,Ogorodnikova2015}. Tritium
retention in tungsten exposed to fusion-relevant plasmas has been shown to
be enhanced markedly once helium and displacement damage introduce
additional trapping sites \cite{Kobayashi2013,Hino2013}, and the resulting
inventory is estimated by integrating the trapping and diffusion terms of
these continuum models over the operating history. The fidelity of that
estimate is limited by the transport coefficients fed into the equations.
Both experiment and simulation show that the hydrogen diffusivity and the
thermal conductivity of tungsten are not single bulk constants: helium
bubbles trap hydrogen and lower its effective
diffusivity by orders of magnitude \cite{Wang2024effD}, grain
boundaries add further trapping sites, while the same
defects scatter electrons and phonons and degrade the conductivity
\cite{Habainy2018,Sina2024}. An accurate component-scale calculation
therefore requires \emph{structure-resolved} coefficients supplied as
position-dependent, and in general anisotropic (tensor), inputs
$\bm{D}(\mathbf{x},T)$ and $\bm{\kappa}(\mathbf{x},T)$, because helium
bubbles and oriented boundaries make the local transport directional
(Fig.~\ref{fig:motivation}).

Atomistic simulation can supply these coefficients: given an atomic
configuration, equilibrium molecular dynamics (MD) yields the phonon
thermal conductivity tensor via Green--Kubo (GK) integration and the
hydrogen diffusion tensor via tracer mean-square displacement (MSD).
However, a converged GK/MSD evaluation of a single $10^4$-atom
microstructure costs hours of CPU time, and the anisotropy of realistic
defected structures demands the full $3\times3$ tensors rather than
scalars. Screening microstructural states along a component lifetime, or
coupling to finite-element models with spatially varying damage, is out of
reach for direct MD.

Machine-learning surrogates offer a way out, and two mature research
threads border on, but do not cover, the present problem. First,
equivariant graph neural networks (GNNs) predict tensorial properties of
crystals, e.g.\ elasticity \cite{Wen2024matten}, dielectric
\cite{Lou2024anisonet} and optical response tensors, directly from the
unit cell; these target \emph{static} properties of \emph{pristine}
database crystals, where the label is a single DFT calculation. Second,
machine-learning interatomic potentials for tungsten--hydrogen now
reach near-DFT fidelity for hydrogen trapping and bubble physics
\cite{Wang2026nep}, complementing established classical potentials for
the full W--H--He system
\cite{Bonny2014}; they make the MD label cheaper per time step
but still require a full MD campaign per microstructure. What is missing
is a surrogate that maps a \emph{defected, disordered microstructure}
directly to its \emph{dynamic transport tensors}, with an explicit and
physically constrained temperature dependence.

This work fills that gap for tungsten containing helium bubbles and grain
boundaries. Its contributions are fourfold:
\begin{enumerate}
\item \textbf{Tensor-valued, temperature-resolved surrogate.} A
message-passing GNN acts on the raw atomic configuration (W and He host
atoms) and returns temperature-independent physical
parameters (an Arrhenius prefactor tensor $D_0$ and activation energy
$E_a$ for hydrogen diffusion, a phonon conductivity tensor
$\kappa_\mathrm{ph}$, and a defect residual resistivity
$\rho_\mathrm{def}$), from which $\Dh(T)$ and $\kappa(T)$ follow
analytically through the Arrhenius and Wiedemann--Franz--Matthiessen
relations at any temperature. The physics constraints guarantee sane
extrapolation in $T$ and reduce the learning problem to 14 parameters.
\item \textbf{Lightweight rotation-equivariant tensor output.} Anisotropy
is encoded by a geometric tensor pooling layer,
$\bm{T}=\sum_{ij}s_{ij}\,\hat{\bm{e}}_{ij}\otimes\hat{\bm{e}}_{ij}$, built
from rotation-invariant edge scores $s_{ij}$ and unit bond vectors
$\hat{\bm{e}}_{ij}$. The output tensor is exactly equivariant under
rotation without spherical-harmonic machinery, keeping the model small
enough to train on hundreds of samples.
\item \textbf{Cost-realistic hybrid teacher strategy.} Labels for 512
microstructures are produced with the Bonny W--H--He EAM potential
\cite{Bonny2014}; the electronic conduction channel, absent in classical
MD, is added by a Wiedemann--Franz--Matthiessen model calibrated to
published irradiation and helium-implantation degradation data
\cite{Habainy2018,Sina2024}; and the whole pipeline is anchored to a
first-principles machine-learning potential (on-the-fly Gaussian-process
FLARE model trained on VASP \cite{Vandermause2020}) through paired
long-time MD calibration runs and a DFT active-learning loop targeting the
most uncertain bubble environments. The measured MLP/EAM diffusivity
ratio (overall geometric mean $0.99$ over 37 paired structures, but
systematically class-resolved, faster in polycrystals and slower in single
crystals) provides a class-dependent correction to the absolute scale.
\item \textbf{Learned trapping physics.} The surrogate reproduces the
hydrogen migration barrier of pristine tungsten
($E_a=0.21$\,eV vs.\ $\sim$0.2\,eV in the literature
\cite{Frauenfelder1969,Heinola2010}) and learns the increase of the
effective barrier in bubble- and GB-containing structures, i.e.\ trapping
emerges from the data rather than being imposed.
\end{enumerate}

The remainder of the paper describes the dataset and teacher calculations
(Sec.~\ref{sec:methods}), validates the rotation equivariance, the
temperature model, the trapping physics and the first-principles anchoring
against experimental and literature data (Sec.~\ref{sec:results}),
demonstrates the surrogate as constitutive input to a coupled
thermal--hydrogen finite-element analysis across scales
(Sec.~\ref{sec:application}), and discusses scope and limitations
(Sec.~\ref{sec:discussion}).

\section{Methods}
\label{sec:methods}

\begin{figure*}[t]
\centering
\includegraphics[width=\linewidth]{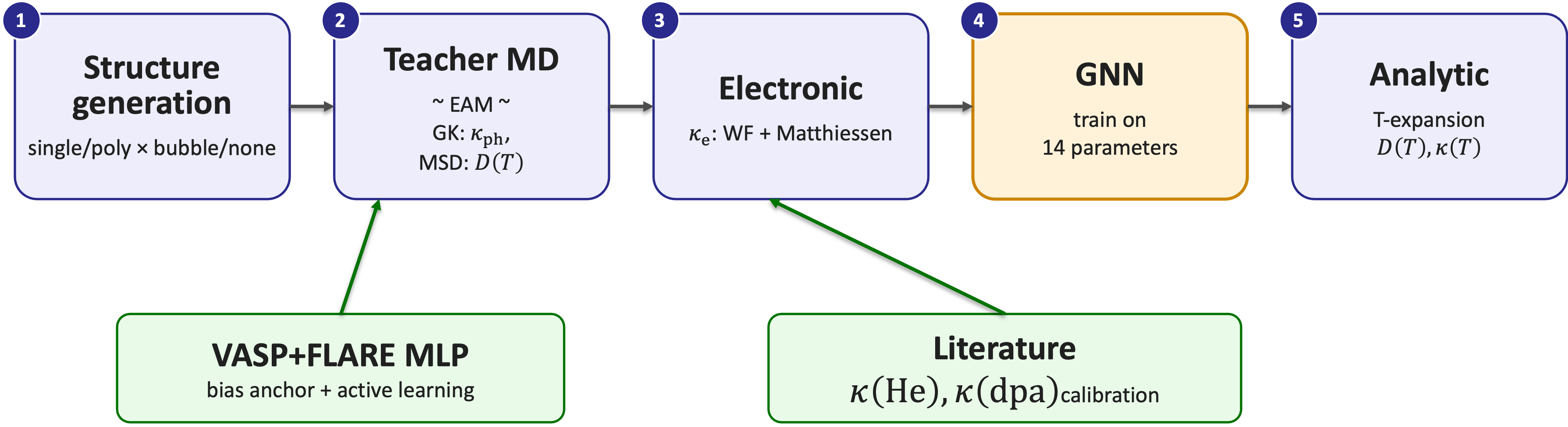}
\caption{Surrogate-construction pipeline. Boxes 1--5 are the forward
workflow; the green data blocks are the two external anchors: an
experimentally calibrated electronic-conduction model
(Sec.~\ref{sec:kappae}) and a first-principles machine-learning potential
used for bias calibration and active learning (Sec.~\ref{sec:mlp}).}
\label{fig:pipeline}
\end{figure*}

The overall workflow is summarised in Fig.~\ref{fig:pipeline}: a
reproducible ensemble of defected microstructures is generated, labelled by
molecular dynamics with an inexpensive potential and an experimentally
calibrated electronic-conduction model, and used to train the GNN; the EAM
teacher and the surrogate are then anchored to a first-principles
machine-learning potential through calibration runs and an active-learning
loop.

\subsection{Microstructure ensemble}
\label{sec:structures}
\begin{figure}[!t]
\centering
\includegraphics[width=0.8\linewidth]{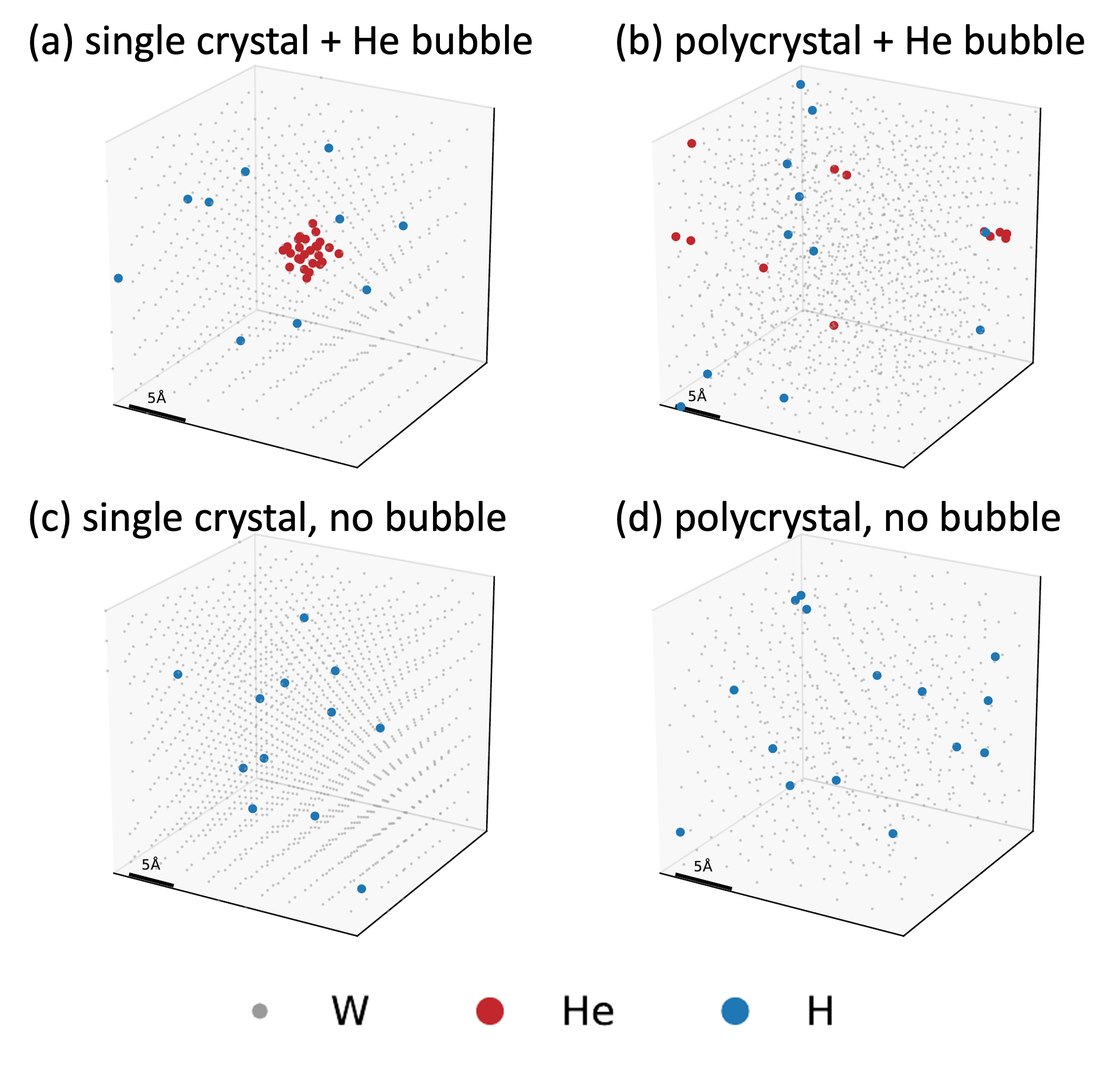}
\caption{Representative held-out test structures, one per class: tungsten
host (grey), helium (red), and the dilute hydrogen tracer (blue), shown in
the periodic simulation cell. (a) A single crystal with a compact helium
bubble; (b) a polycrystal in which helium decorates grain boundaries;
(c,d) the corresponding bubble-free single crystal and polycrystal.
These configurations are the GNN input (W and
He only); the hydrogen tracer is the diffusing species whose transport is
predicted.}
\label{fig:samples}
\end{figure}

We generate 512 periodic tungsten microstructures spanning four classes:
single crystal and Voronoi polycrystal, each with and without helium
bubbles. Structure parameters (cell size 6--10\,$a_0$ blocks,
bubble radius 0.12--0.26 of the box edge, He-to-vacancy ratio 0.8--1.6,
one or two bubbles, random grain orientations) are sampled
deterministically from the structure index so that the ensemble is
reproducible. Polycrystals are approximated by a periodic Voronoi
tessellation: $n$ grain centres ($n=2$--$4$) are placed at random in the
cubic cell, each grain is filled with an independently, randomly oriented
bcc crystal, and every atom is assigned to its nearest centre under the
minimum-image convention. Atoms that would overlap across a grain boundary
are removed whenever a kept neighbour lies closer than
$r_\mathrm{ov}=2.0$\,\AA{} ($0.63\,a_0$, well below the bcc
nearest-neighbour distance $0.87\,a_0$); this overlap removal defines the
grain-boundary open volume, and no additional vacuum gap is inserted.
Hydrogen is inserted as a dilute tracer at interstitial and
open-volume sites (bubble peripheries and the grain-boundary open volume,
with host separations of $1.7$--$2.7$\,\AA{} for helium and
$1.3$--$2.2$\,\AA{} for the hydrogen tracer). Concretely, the cubic cell
edge is $6$--$10\,a_0$ ($\approx1.9$--$3.2$\,nm), the single-crystal bubble
radius is $0.12$--$0.26$ of the box edge, the He-to-vacancy ratio is
$0.8$--$1.6$, polycrystals have 2--4 grains, and 8--19 hydrogen tracers are
inserted; a trapping-focused extension adds denser and over-pressurised
bubbles, scattered vacancies and vacancy--hydrogen complexes. All
structures are relaxed (conjugate-gradient minimisation followed by a short
NVT equilibration) before label generation.
Representative members of the four classes, drawn from the held-out test
set, are shown in Fig.~\ref{fig:samples}; they range from 400 to 2000
atoms and differ markedly in the spatial arrangement of helium (a compact
bubble in single crystals, boundary-decorating clusters in polycrystals)
and in the trapping environment seen by the hydrogen tracer.

\subsection{Teacher molecular dynamics}
\label{sec:teacher}
Labels are computed with LAMMPS using the Bonny W--H--He EAM2 potential
\cite{Bonny2014}, chosen for its tractable cost and its qualitatively
correct He clustering and H trapping behaviour; its quantitative bias is
measured against a first-principles MLP in Sec.~\ref{sec:mlp}.

\paragraph{Hydrogen diffusion} The hydrogen diffusion tensor is obtained
from a tracer molecular-dynamics run in the canonical (NVT) ensemble. The
relaxed configuration is integrated with a 1\,fs time step under
three-dimensional periodic boundary conditions; the tungsten and helium
atoms form the fixed host, and hydrogen (introduced as a dilute tracer, so
that H--H interactions are negligible and each H atom samples the local
trapping landscape independently) is the diffusing species. Velocities are
initialised from a Gaussian distribution at the target temperature and the
temperature is controlled by a Nos\'e--Hoover thermostat with a 0.1\,ps
damping constant. After $2\times10^4$ steps (20\,ps) of equilibration, the
unwrapped hydrogen coordinates are recorded every 200 steps (0.2\,ps) over
a $3\times10^5$-step (300\,ps) production run, yielding $\sim$1500 frames.
The full mean-square-displacement (MSD) tensor is accumulated with multiple
time origins,
\begin{equation}
M_{\alpha\beta}(\tau)=\big\langle
\Delta r_\alpha(\tau)\,\Delta r_\beta(\tau)\big\rangle_{t_0,\,\mathrm{H}},
\qquad
D_{\alpha\beta}=\tfrac12\,\frac{\mathrm{d}M_{\alpha\beta}}{\mathrm{d}\tau},
\label{eq:msd}
\end{equation}
where $\alpha,\beta\in\{x,y,z\}$ label the Cartesian components of the
tracer displacement $\Delta\bm{r}(\tau)$ (so that $M_{\alpha\beta}$ and
$D_{\alpha\beta}$ are the components of the $3\times3$ MSD and diffusion
tensors), the average runs over time origins $t_0$ and all tracer atoms, and
the slope is taken by linear regression over the lag window spanning
25--100\% of the accessible range (the ballistic short-time part is
excluded); the resulting tensor is symmetrised and converted to
$\mathrm{m^2\,s^{-1}}$. Repeating this at four temperatures
(700, 900, 1100, 1400\,K) and fitting the isotropic diffusivity
$D_\mathrm{iso}(T)=\tfrac13\,\mathrm{tr}\,\bm{D}$ to an Arrhenius law
$\ln D_\mathrm{iso}=\ln D_{0,\mathrm{iso}}-E_a/k_BT$ gives a common
activation energy $E_a$; the prefactor tensor is then recovered as
$\bm{D}_0=\langle\bm{D}(T)\,e^{E_a/k_BT}\rangle_T$. The pair $(\bm{D}_0,E_a)$
constitutes the diffusion labels. In bubble- and GB-containing structures
the tracer spends part of the run trapped, which lowers $D_\mathrm{iso}(T)$
and raises the fitted $E_a$; both effects are learnt by the surrogate.

\paragraph{Phonon thermal conductivity} $\kappa_\mathrm{ph}$ at 300\,K is
computed by Green--Kubo integration of the heat-flux autocorrelation,
$\kappa_{\alpha\beta}=\frac{1}{k_BT^2V}\int_0^\infty\langle
J_\alpha(0)J_\beta(t)\rangle\,dt$, averaged over three independent
velocity seeds ($4\times10^5$ steps each) to suppress GK noise; the tensor
is symmetrised.

\subsection{Electronic thermal conduction}
\label{sec:kappae}
In tungsten $\sim$90\% of heat is carried by electrons, which classical MD
cannot describe. We add an electronic channel through the
Wiedemann--Franz law with a Matthiessen decomposition of the resistivity,
\begin{equation}
\kappa_e(T)=\frac{L_\mathrm{eff}\,T}{\rho(T)},\qquad
\rho(T)=\rho_W(T)+A_\mathrm{dis}f_\mathrm{dis}
+A_\mathrm{He}c_\mathrm{He}+A_\mathrm{H}c_\mathrm{H},
\label{eq:wf}
\end{equation}
where $\rho_W(T)$ is the recommended analytical electrical resistivity of
tungsten (White--Minges fit)~\cite{Tolias2017}, which is superlinear at high
temperature so that $\kappa_e$ decreases with temperature as physically
observed, and $L_\mathrm{eff}=2.97\times10^{-8}\,\mathrm{W\,\Omega\,K^{-2}}$
is an effective Lorenz number (slightly above the ideal Sommerfeld value,
the excess absorbing the small lattice contribution) calibrated so that the
pristine conductivity reproduces the recommended tungsten values to within
$\sim$5\% over 300--1400\,K. In Eq.~\eqref{eq:wf}, $f_\mathrm{dis}$ is the fraction of W atoms whose coordination
deviates from bcc (a GB/interface/point-defect measure computed from the
configuration), $c_\mathrm{He}$, $c_\mathrm{H}$ are atomic fractions, and
$\rho$ is capped at the Ioffe--Regel saturation value
$\rho_\mathrm{sat}=1.5\,\mu\Omega$\,m to prevent unphysical extrapolation
in bubble-dense regions. The coefficients are calibrated to published
room-temperature degradation measurements: helium-implanted tungsten
(transient grating spectroscopy at 280 and 3100 appm~\cite{Hofmann2015};
the calibration adopts slightly conservative representative ratios
$\kappa/\kappa_0=0.70$ and $0.45$, cf.\ the as-measured room-temperature
diffusivity ratios of ${\approx}0.78$ and ${\approx}0.53$) and the
$\sim$0.5 saturation of neutron/proton-irradiated tungsten above
0.5\,dpa \cite{Habainy2018}. Because the phonon and electron channels
cannot be separated in the experiments, the calibration is bracketed by
two scenarios (phonon conductivity unchanged / degraded proportionally to
the electronic channel) and the geometric mean is adopted:
$A_\mathrm{He}=4.9\times10^{-5}$, $A_\mathrm{dis}=7.2\times10^{-6}$,
$A_\mathrm{H}=1.6\times10^{-5}\;\Omega$\,m per atomic fraction
(the hydrogen coefficient is not constrained by these hydrogen-free
experiments and is set by the assumption
$A_\mathrm{H}=A_\mathrm{He}/3$). The calibrated model, applied without further adjustment,
reproduces the independent degradation data of proton/spallation-irradiated
tungsten of Sina \emph{et al.} \cite{Sina2024}
($\kappa/\kappa_0=0.45$--0.46 measured at 50\,$^\circ$C by laser-flash
analysis, at 410--1225 appm of co-produced helium on top of saturated
displacement damage; predicted 0.46--0.52). We note that Sina \emph{et al.}
attribute the measured degradation primarily to displacement damage rather
than to helium itself; this comparison therefore validates the \emph{total}
predicted degradation of the co-irradiated material, not the partition
between the displacement and helium channels. The total label is
$\kappa=\kappa_\mathrm{ph}+\kappa_e\bm{I}$. Two evaluation regimes of
Eq.~\eqref{eq:wf} must be distinguished. The \emph{label} is evaluated with
the full composition of the simulation cell, including the hydrogen tracer
($c_\mathrm{H}=0.4$--$5$\,at.\%), whose contribution
$A_\mathrm{H}c_\mathrm{H}$ is comparable to or larger than
$\rho_W(300\,\mathrm{K})$; together with the helium and disorder terms and
the saturation cap, this places most defected labels far below pristine
tungsten at room temperature (Sec.~\ref{sec:data}). All
\emph{application-side} evaluations of $\kappa_e$
(Sec.~\ref{sec:kappacal} onwards) are instead performed in the
hydrogen-dilute limit $c_\mathrm{H}=0$, appropriate for a component in
which hydrogen is a dilute interstitial.

\subsection{Graph neural network with geometric tensor pooling}
\label{sec:gnn}
\begin{figure}[t]
\centering
\includegraphics[width=\linewidth]{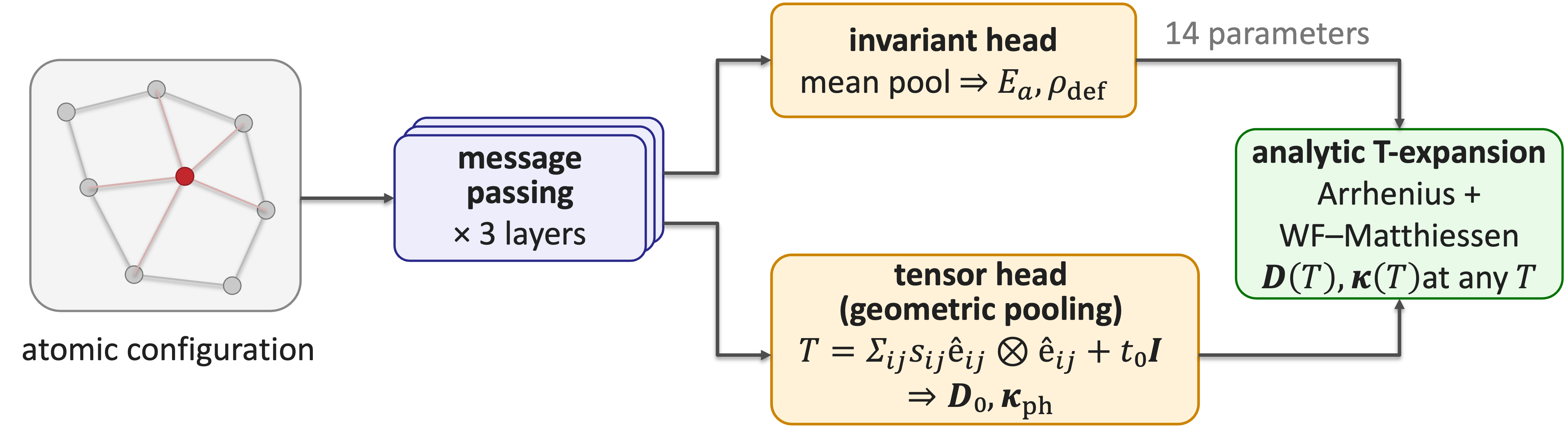}
\caption{GNN architecture. Shared message-passing layers feed two output
heads: a rotation-invariant head for the scalars $E_a$, $\rho_\mathrm{def}$
and a rotation-equivariant geometric-tensor-pooling head
(Eq.~\ref{eq:pool}) for the tensors $\bm{D}_0$, $\bm{\kappa}_\mathrm{ph}$.
The 14 temperature-independent parameters are expanded analytically to any
temperature through the Arrhenius and Wiedemann--Franz--Matthiessen
relations (Eq.~\ref{eq:thead}).}
\label{fig:gnn}
\end{figure}

The input graph contains the host atoms (W, He; the dilute H tracer is
the predicted species and is excluded from the input) with
periodic-boundary edges within a cutoff $r_c=4$\,\AA. Node features are
species embeddings; edge features are interatomic distances expanded in
radial basis functions. After three message-passing layers the model
branches into (i) an invariant head producing scalars ($E_a$,
$\rho_\mathrm{def}$) by mean pooling, and (ii) a tensor head producing
$D_0$ and $\kappa_\mathrm{ph}$ via geometric tensor pooling
\begin{equation}
\bm{T}=\sum_{(ij)}s_{ij}\,
\hat{\bm{e}}_{ij}\otimes\hat{\bm{e}}_{ij}+t_0\bm{I},
\label{eq:pool}
\end{equation}
in which the edge scores $s_{ij}$ and the isotropic offset $t_0$ are
rotation-invariant functions of the learned features and
$\hat{\bm{e}}_{ij}$ is the unit bond vector. Since rotations act only on
$\hat{\bm{e}}_{ij}$, $\bm{T}$ transforms exactly as a rank-2 tensor and is
symmetric by construction. The network thus predicts the 14
temperature-independent parameters
$\{D_0(6),E_a,\kappa_\mathrm{ph}(6),\rho_\mathrm{def}\}$ and inference at
temperature $T$ evaluates
\begin{equation}
\Dh(T)=D_0\,e^{-E_a/k_BT},\qquad
\kappa(T)=\kappa_\mathrm{ph}\frac{T_\mathrm{ref}}{T}
+\frac{L_\mathrm{eff}\,T}{\rho_W(T)+\rho_\mathrm{def}}\bm{I}.
\label{eq:thead}
\end{equation}
The architecture is shown in Fig.~\ref{fig:gnn}.

\subsection{Training protocol}
\label{sec:training}
The 14-component target vector (six $D_0$ and six $\kappa_\mathrm{ph}$
tensor entries, plus the scalars $E_a$ and $\rho_\mathrm{def}$) is
regressed with an $L_1$ loss in which the two scalars carry twice the
weight of the tensor components, reflecting their importance for the
temperature expansion. The tensor labels are transformed under the same
random rotations as the input during augmentation, so the model learns the
covariant mapping; the heavy-tailed $D_0$ is trained through the
$\log_{10}$ parameterisation of the constitutive closure with z-score
normalisation computed on the training split only. Optimisation uses Adam (learning rate
$3\times10^{-4}$), gradient-norm clipping and cosine annealing, which we
found necessary to stabilise small-sample GNN training. All reported
metrics are means $\pm$ standard deviation over five independent
train/validation splits and seeds; single-split comparisons of the two
architectures were found to fluctuate by more than their difference.

\subsection{First-principles anchoring and active learning}
\label{sec:mlp}
To quantify the EAM teacher's bias we built a first-principles
machine-learning potential for W--H--He with the FLARE
Gaussian-process framework \cite{Vandermause2020} trained on-the-fly on
VASP (PBE, ENCUT 600\,eV, $k$-spacing 0.2\,\AA$^{-1}$, W\_sv
pseudopotential) over a seed library of bulk, GB, vacancy, and He-cluster
structures. Because the MLP costs $\sim$3600$\times$ more per MD step than
the EAM, it is used for anchoring rather than label production: paired
long-time diffusion runs over 37 structures give an MLP/EAM diffusivity
ratio whose overall geometric mean is $0.99$ but which is systematically
class-resolved: polycrystals $2.7$ (with bubbles) and $1.7$ (without),
single crystals $0.57$ and $0.40$, i.e.\ the embedded-atom model
over-traps at grain boundaries and under-traps at isolated vacancies.
These class-dependent factors are applied to the EAM diffusion labels
before the final retraining (Sec.~\ref{sec:anchor}). A dedicated DFT
campaign (59 seed structures, of which 39 sample VH$_{1\text{--}6}$,
di-vacancy--H, He--vacancy--H complexes and small bubbles with hydrogen)
was subsequently collected to densify exactly these environments. A direct
static validation shows that the production anchor MLP already reproduces
the mono-vacancy hydrogen binding energy, $E_b(\mathrm{V\!-\!H})=1.15$\,eV
against the DFT range 1.0--1.4\,eV spanned by the zero-point-corrected
sequential binding energies of the first five hydrogen atoms to the
monovacancy \cite{HeinolaVH2010}, confirming that the vacancy-trapping deficiency
resides in the embedded-atom teacher, not in the anchor, and is corrected
by the class-resolved calibration above. A refit of the MLP on the
trap-augmented DFT set was subsequently obtained with a gradient-free
hyperparameter optimisation (the gradient-based likelihood optimiser is
numerically fragile at this dataset size, and a refit with default
hyperparameters over-binds, $E_b=3.1$\,eV). The refit reproduces the
static DFT benchmarks ($E_b(\mathrm{V\!-\!H})=1.06$\,eV; a
tetrahedral-to-tetrahedral migration barrier of $0.20$\,eV against
$\sim$0.2\,eV from DFT \cite{Heinola2010}), but paired
multi-temperature diffusion runs show its absolute hydrogen diffusivity
in pristine tungsten to exceed the experimental gas-loading scale
\cite{Frauenfelder1969} by a factor of 2--6. The production anchor,
whose paired calibration is consistent with the experimentally
validated teacher labels, is therefore retained for the released
model, and the refit is deliberately not deployed.

\section{Results}
\label{sec:results}
Unless stated otherwise, all predictions reported below are made with the
final surrogate, i.e.\ the model retrained on the first-principles
anchor-corrected diffusion labels (Sec.~\ref{sec:anchor}); the anchoring
procedure and its class-resolved corrections are described there before the
literature validation that relies on them.

\subsection{Dataset statistics and learnability}
\label{sec:data}
The teacher set comprises 635 relaxed microstructures: 512 spanning the four
base classes (single crystal / polycrystal $\times$ helium bubble
present / absent) and an additional 123 trapping-focused configurations with
denser and over-pressurised bubbles, scattered vacancies and grain-boundary
helium (Sec.~\ref{sec:teacher}). Across the set the isotropic hydrogen
diffusivity $D_0^\mathrm{iso}$ spans roughly two decades
($\sim\!10^{-9}$--$10^{-7}$\,m$^2$\,s$^{-1}$), fast in ordered single
crystals and slow in bubble- and boundary-rich structures, while the
electronically calibrated conductivity label, evaluated at 300\,K with the
full cell composition, spans $4$--$136$\,W\,m$^{-1}$K$^{-1}$ (median
$17$): the hydrogen-tracer, helium and disorder resistivity terms and the
Ioffe--Regel cap pull defected cells far below the pristine value
(Sec.~\ref{sec:kappae}). Helium content correlates
with the defect residual resistivity ($r=+0.29$) and, after the trapping
campaign, with the migration barrier (open-volume fraction vs.\ $E_a$:
$r=+0.43$, versus $+0.22$ for the base set), confirming a learnable
structural signal in both channels.

\subsection{Rotation equivariance}
\label{sec:equivariance}
Because a divertor microstructure is sampled as arbitrarily oriented
sub-blocks of a larger specimen, the surrogate must be orientation
consistent. Evaluated on single-crystal blocks cut at eight random
orientations, the retrained non-periodic GNN predicts an activation energy
with a coefficient of variation of only $0.2$--$0.3\%$ and $D_0^\mathrm{iso}$
to within $\sim\!1\%$: scalar invariants are effectively rotation invariant,
while the tensor outputs rotate covariantly through the geometric pooling
layer (Sec.~\ref{sec:gnn}). This property is what allows the continuum
fields of Sec.~\ref{sec:application} to be assembled from blocks of
arbitrary lattice orientation.

\subsection{Temperature dependence and trapping physics}
\label{sec:trapping}
The predicted migration barrier has a median of $0.21$\,eV, consistent with
the low end of reported tungsten hydrogen migration energies
\cite{Frauenfelder1969,Heinola2010} for the classical potential used.
The trapping-focused MD campaign resolves the defect dependence: relative to
the base set, helium fraction versus $E_a$ turns from $r=-0.08$ to
$r=+0.28$, and the effective diffusivity at 600\,K falls to $\sim\!45\%$ of
the pristine value. The signal is strongest for grain boundaries
(class-median $E_a=0.23$\,eV, $D(600\,\mathrm{K})$ reduced $\sim\!12\times$)
and for dense bubbles, whereas isolated vacancies show only a weak up-shift,
a limitation traced to the embedded-atom description of H--vacancy binding
(Sec.~\ref{sec:discussion}). Thus grain-boundary and bubble trapping, the
dominant effective-diffusivity reducers in irradiated tungsten, are
captured, and the temperature-independent parameterisation lets the
Arrhenius expansion reproduce the diffusivity at arbitrary temperature from a
single inference.

\subsection{Electronic conductivity calibration}
\label{sec:kappacal}
The conductivity is dominated by the electronic channel; we therefore
evaluate $\kappa_e$ from a Wiedemann--Franz--Matthiessen model
(Sec.~\ref{sec:kappae}) using the recommended analytical
temperature-dependent electrical resistivity of tungsten $\rho_W(T)$
(White--Minges fit)~\cite{Tolias2017} with an effective Lorenz number
$L_\mathrm{eff}=2.97\times10^{-8}\,\mathrm{W\,\Omega\,K^{-2}}$ (slightly
above the ideal Sommerfeld value, absorbing the small lattice contribution),
and add the GNN phonon tensor (a few percent of the total). Because
$\rho_W(T)$ is superlinear at elevated temperature, $\kappa_e$ decreases with
temperature as physically expected, reproducing the recommended
pure-tungsten conductivity to within $\sim$5\% between 300 and 1400\,K. For
helium \emph{bubbles}, which
are voids rather than dissolved scatterers, the reduction is modelled as an
effective-medium (Maxwell) porosity in the local void fraction $\phi$,
$\kappa=\big[(1-\phi)/(1+\phi/2)\big]\,[\kappa_{ph}+\kappa_e]$. This prefactor
is the classical Maxwell--Eucken result for a dilute, non-interacting
dispersion of spherical inclusions in the limit of vanishing inclusion
conductivity ($\kappa_\mathrm{void}\!\to\!0$), obtained by setting the
inclusion conductivity to zero in the Maxwell mixing rule; for a two-phase
solid it coincides with the Hashin--Shtrikman upper bound with the conducting
tungsten as the continuous phase~\cite{Hashin1962}, i.e.\ the
``internal-porosity'' limit appropriate for isolated gas-filled bubbles
suspended in the matrix~\cite{Carson2005}. This yields a
pristine value $\kappa(800\,\mathrm{K})\simeq129$\,W\,m$^{-1}$K$^{-1}$
decreasing to $110$ and $94$\,W\,m$^{-1}$K$^{-1}$ at $\phi=0.10$
and $0.20$ (Fig.~\ref{fig:closure}); grain boundaries reduce the phonon
term only, a negligible correction for electron-dominated tungsten. Directly
regressing the residual resistivity with the GNN instead collapsed the
pristine conductivity unphysically, because the analytically injected
resistivity labels saturate at the Ioffe--Regel limit for the majority of
defected structures; the analytic electronic model is therefore retained for
$\kappa_e$ while the GNN supplies the phonon tensor and the full diffusion
parameterisation.

\subsection{First-principles anchoring and active learning}
\label{sec:anchor}
The bulk teacher labels use the fast embedded-atom potential; a
first-principles machine-learning potential (VASP+FLARE SGP) anchors the
absolute scale. Paired MD on 37 identical structures gives a diffusivity
ratio $D_\mathrm{MLP}/D_\mathrm{EAM}$ whose overall geometric mean is close
to unity ($0.99$) but which is systematically \emph{class-resolved}: for
polycrystalline structures the machine-learning potential predicts faster
hydrogen transport (geometric-median ratio $2.7$ with bubbles, $1.7$
without), i.e.\ the embedded-atom model over-traps at grain boundaries,
whereas for single crystals it predicts slower transport ($0.57$ with
bubbles, $0.40$ without), i.e.\ the embedded-atom model under-traps at
isolated vacancies. Applying these class-resolved corrections to the EAM
diffusion labels and retraining lowers the pristine single-crystal
pre-exponential $D_0$ by a factor of $\sim\!2.4$ (bringing it onto the
first-principles scale) while leaving the migration energy essentially
unchanged (median $E_a=0.21$\,eV), thereby sharpening the diffusivity
contrast between ordered and defected regions and strengthening the
vacancy-trapping channel that the embedded-atom model underestimates.
The SGP uncertainty ($c_\mathrm{unc}$, here $0.1$--$0.6$) simultaneously
flags structures near the edge of the potential's training domain; a
59-structure DFT campaign densifying H--vacancy and He--vacancy
environments has been collected on this basis, and a static validation
shows the production anchor already reproduces the mono-vacancy hydrogen
binding energy ($1.15$\,eV vs.\ the 1.0--1.4\,eV spread of
zero-point-corrected sequential DFT binding energies
\cite{HeinolaVH2010}); a trap-augmented refit of the anchor was
evaluated but deliberately not adopted, because its dynamic
diffusivities overshoot the experimental scale (Sec.~\ref{sec:mlp}).

\subsection{Validation against experimental and literature data}
\label{sec:valid}
\begin{figure}[!t]
\centering
\includegraphics[width=\linewidth]{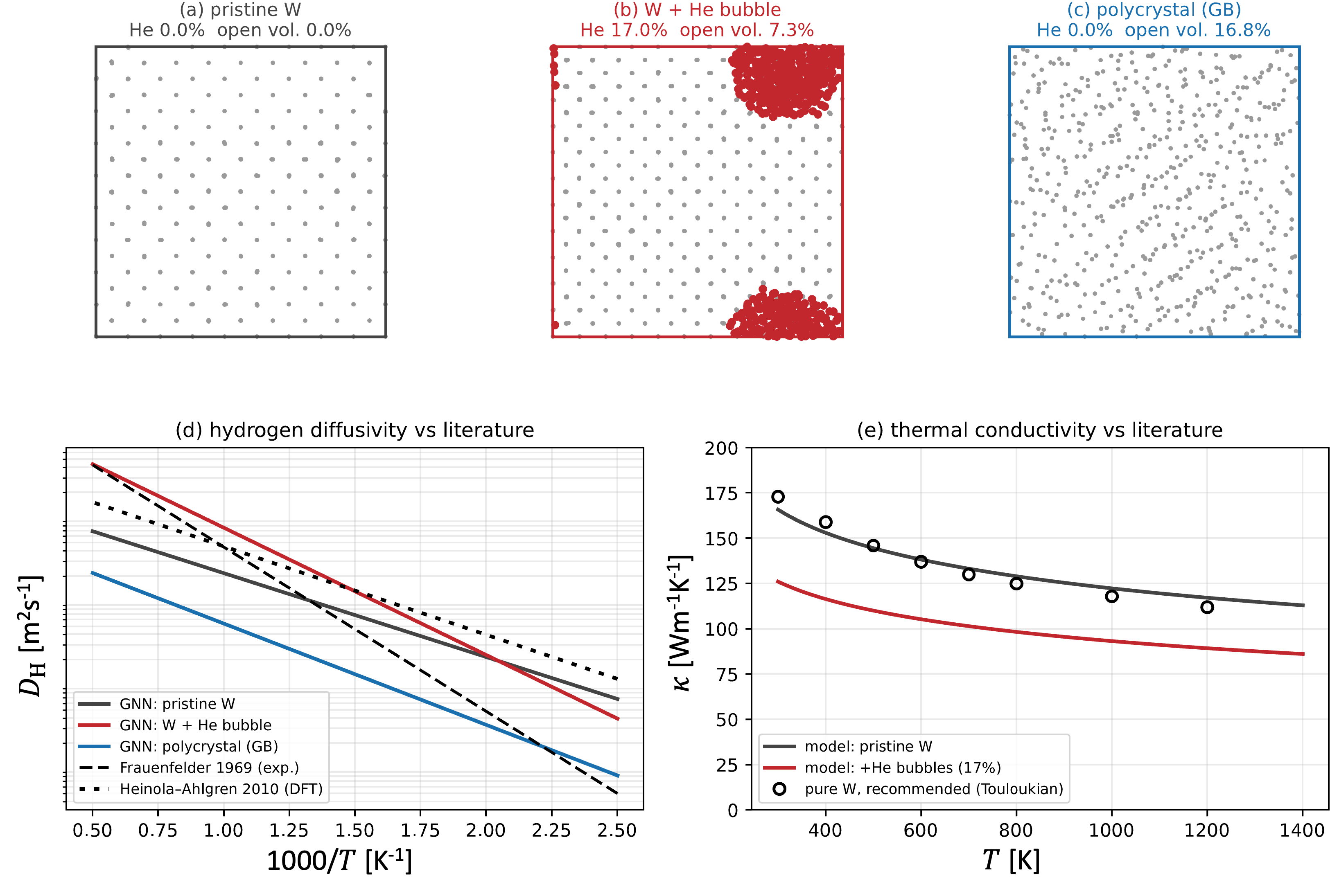}
\caption{Validation against literature. (a--c) Representative teacher
microstructures ($xy$ projections; grey: W, red: He): pristine single
crystal, single crystal with pressurised helium bubbles, and a polycrystal
with grain-boundary open volume. (d) Predicted hydrogen diffusivity for
the three structures compared with the gas-loading experiment of
Frauenfelder \cite{Frauenfelder1969} and the DFT Arrhenius law of
Heinola--Ahlgren \cite{Heinola2010}. (e) Predicted thermal conductivity
compared with recommended pure-tungsten values \cite{Touloukian1970}.}
\label{fig:valid}
\end{figure}

Figure~\ref{fig:valid} confronts the surrogate with experimental and
first-principles literature for both transport channels, alongside the
atomic configurations on which the predictions are made. For pristine
tungsten the predicted hydrogen diffusivity runs parallel to, and within a
factor of $\sim$2 of, the DFT-based Arrhenius law of Heinola and Ahlgren
($D_0=5.2\times10^{-8}$\,m$^2$\,s$^{-1}$, $E_a=0.21$\,eV
\cite{Heinola2010}), and crosses the high-temperature gas-loading
measurement of Frauenfelder \cite{Frauenfelder1969} within its
experimental temperature window. A dense pressurised-bubble structure
(17\% He) raises the predicted barrier to $0.30$\,eV, so its diffusivity
falls below the pristine curve at low temperature (trapping), while
slightly exceeding it in the high-temperature de-trapping regime; this
single representative grain-boundary structure is uniformly slower by a
factor of 3--4 (the larger $\sim$12$\times$ reduction quoted in
Sec.~\ref{sec:trapping} is the class median over all grain-boundary
structures at 600\,K). The
predicted conductivity of pristine tungsten reproduces the recommended
pure-tungsten values \cite{Touloukian1970} to within $\sim$5\% across the
whole 300--1400\,K range, including the correct decreasing temperature
dependence, because the electronic channel uses the recommended
temperature-dependent resistivity $\rho_W(T)$ rather than a linear
approximation (Sec.~\ref{sec:kappacal}). Helium bubbles reduce the predicted
conductivity by the Maxwell porosity factor (Sec.~\ref{sec:kappacal}).

\subsection{Inference cost}
\label{sec:cost}
A converged Green--Kubo/MSD evaluation of one $\sim\!10^3$-atom cell costs
hours of MD, whereas the trained surrogate returns both $3\times3$ tensors in
milliseconds per block, a speed-up of order $10^6$ that makes it feasible to
populate a continuum field with microstructure-resolved coefficients and to
close the loop into a component-scale solver (Sec.~\ref{sec:application}).

\section{Application: coupled thermal--hydrogen transport}
\label{sec:application}
To demonstrate that the surrogate delivers usable continuum input we solve
the coupled heat and hydrogen transport of a helium-bubble polycrystalline
tungsten slab in two steps that mirror the flow of information across
scales: (i)~an explicit relaxed nanoscale volume is evaluated block by
block \emph{directly} by the GNN, and the resulting block statistics are
distilled into a smooth constitutive closure that also supplies the
treatment of the out-of-domain bubble-interior blocks
(Sec.~\ref{sec:nanofield}); (ii)~the closure is carried to the
component scale, where no atomistic structure exists, to drive a coupled
millimetre-scale monoblock analysis (Sec.~\ref{sec:component}).
Two-dimensional linear ($P_1$) finite-element solvers for the steady
anisotropic non-linear heat equation
$\nabla\!\cdot(\bm{\kappa}(T)\nabla T)=0$ (Picard-linearised) and the steady
hydrogen transport $\nabla\!\cdot(\bm{D}\nabla C)=0$ were verified against
analytic uniform-coefficient solutions to machine precision and against a
temperature-dependent-$\kappa$ conduction problem
($<\!10^{-5}$ relative error).

\subsection{Nanoscale fields: direct block-wise evaluation and constitutive closure}
\label{sec:nanofield}
\begin{figure}[!tbp]
\centering
\includegraphics[width=\linewidth]{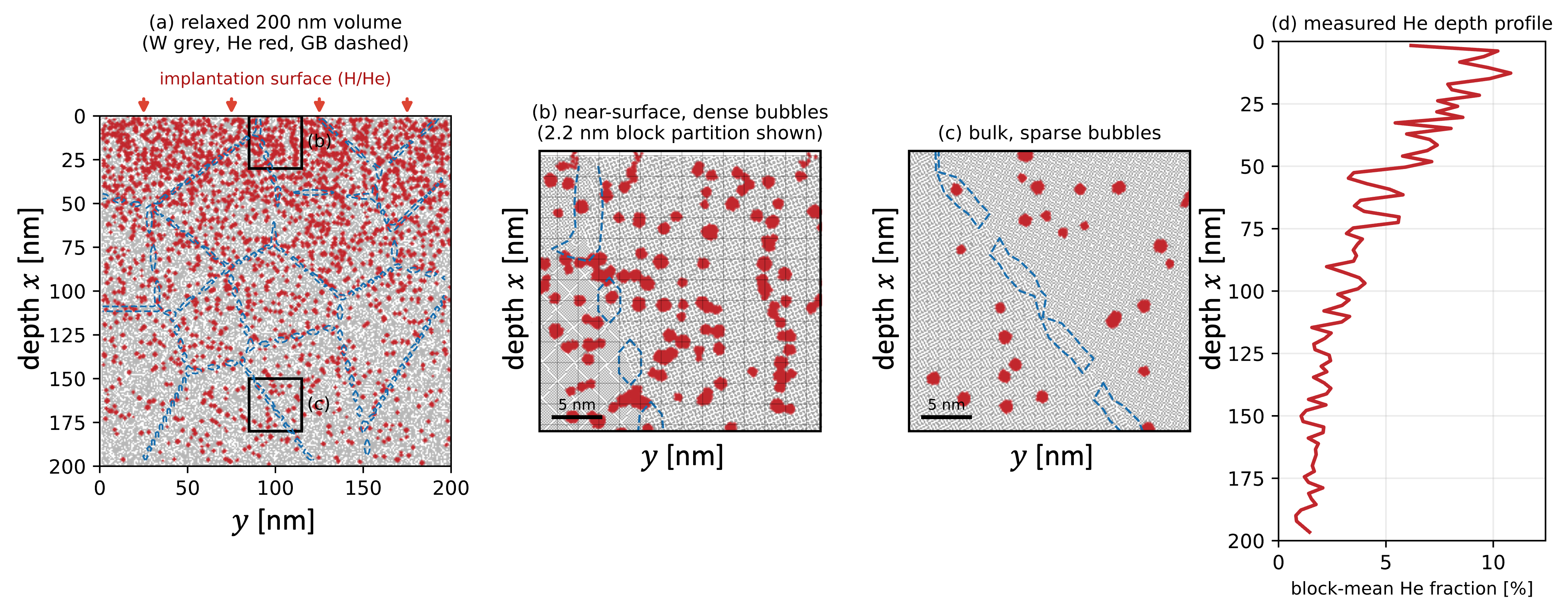}
\caption{Setup of the nanoscale calculation. (a) The relaxed
$200\times200\times2.2$\,nm atomistic volume ($5.6\times10^6$ atoms; W grey,
He red; a $\sim$1\% subsample of atoms is shown), with the grain-boundary
network overlaid (blue dashed, the same contours as in
Fig.~\ref{fig:field}) and the implantation surface on top; boxes mark the
clips of panels (b,c). (b) Helium-rich near-surface region (depth
0--30\,nm, dense bubble population); the thin grid shows the 2.2\,nm block
partition on which the GNN is evaluated. (c) Sparse bulk (depth
150--180\,nm). Grain boundaries between columnar grains of different
in-plane orientation are visible as moir\'e contrast and coincide with the
dashed network. (d) Block-mean helium fraction versus depth, measured from
the volume.}
\label{fig:setup200}
\end{figure}

The nanoscale computational volume is summarised in
Fig.~\ref{fig:setup200}: a $200\times200\times2.2$\,nm
quasi-two-dimensional atomistic volume is generated explicitly
($5.6\times10^6$ atoms: sixteen columnar grains with random in-plane
orientations and $2.2\times10^5$ helium atoms in $\sim$0.7\,nm bubbles whose
density decays from the implantation surface over $\sim$60\,nm), and is
relaxed with the same
molecular-dynamics protocol used for the teacher data (FIRE minimisation, a
5\,ps anneal at 600\,K and a final minimisation with the same interatomic
potential~\cite{Bonny2014}). The relaxed structure is partitioned into
2.2\,nm blocks ($89\times89$) and each block is evaluated \emph{directly} by
the GNN on its non-periodic graph, yielding the local $D_0$, $E_a$ and
phonon tensor; the electronic channel and the bubble porosity are added
analytically as in Sec.~\ref{sec:kappacal}, with the Maxwell factor
evaluated in the void fraction measured per block from the atomistic
configuration. The direct prediction is retained for every block whose
helium fraction lies within the training domain of the model
($f\le0.13$, $88.5\%$ of blocks).

\begin{figure}[!tbp]
\centering
\includegraphics[width=0.78\linewidth]{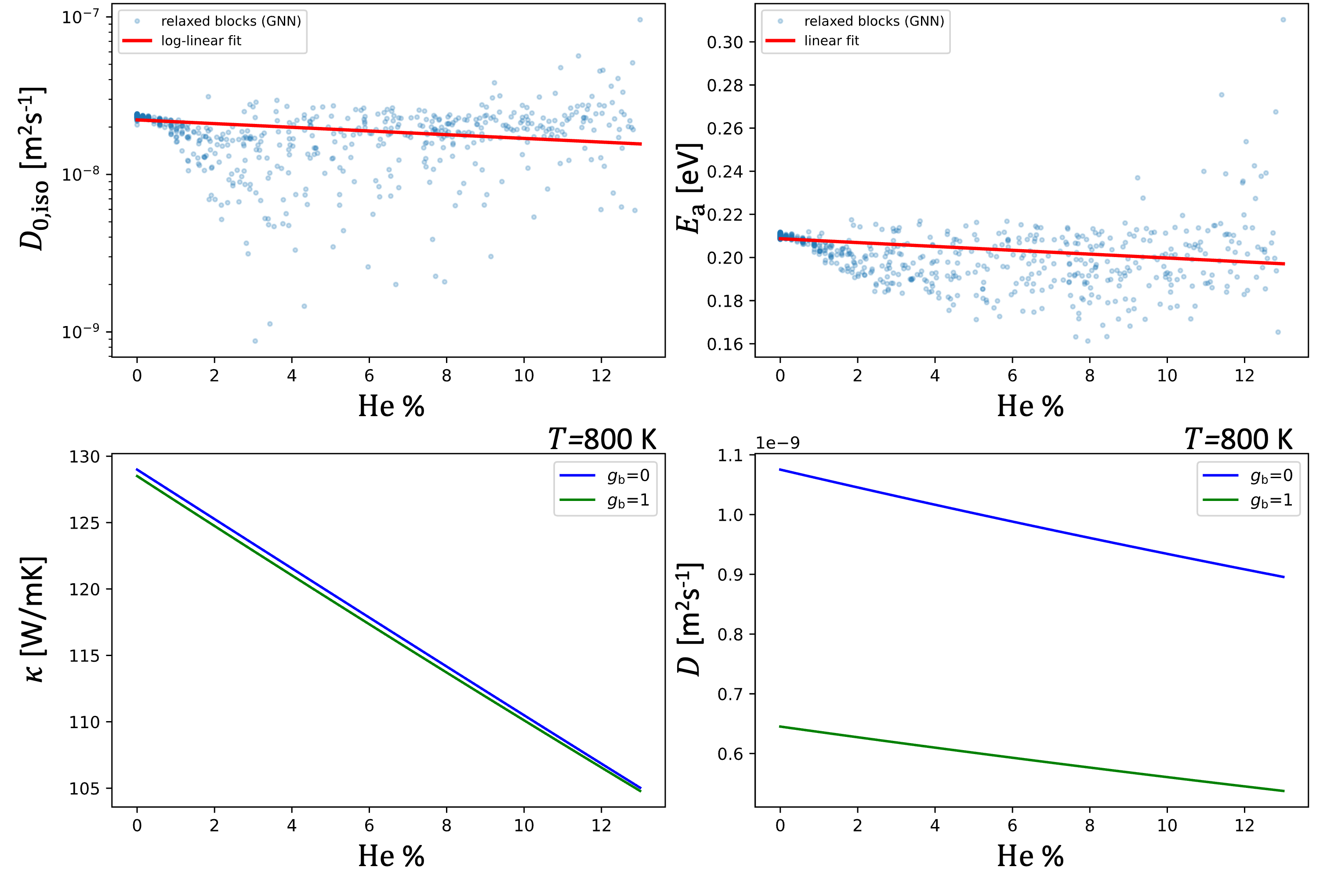}
\caption{Constitutive closure fitted to the direct GNN evaluations of the
2.2\,nm blocks of the relaxed volume of Sec.~\ref{sec:nanofield}
(grain-interior blocks within the training domain). Top: per-block $D_0$
and $E_a$ versus helium fraction, with the log-linear and linear fits used
by the closure (red). Bottom: the resulting $\kappa$ and $D_\mathrm{H}$ at
800\,K without ($g_\mathrm{b}=0$, grain interior) and with
($g_\mathrm{b}=1$, fully on a boundary) grain boundaries, where
$g_\mathrm{b}$ is the grain-boundary fraction of a continuum cell defined
in the text. Helium lowers $\kappa$ through
porosity, while its direct effect on $D_\mathrm{H}$ is mild at this
temperature; grain boundaries are the dominant $D_\mathrm{H}$ reducer. The
closure substitutes for bubble-interior blocks outside the training domain
at the nanoscale and serves as the constitutive law at the component scale
(see text).}
\label{fig:closure}
\end{figure}

\begin{figure}[!tbp]
\centering
\includegraphics[width=0.85\linewidth]{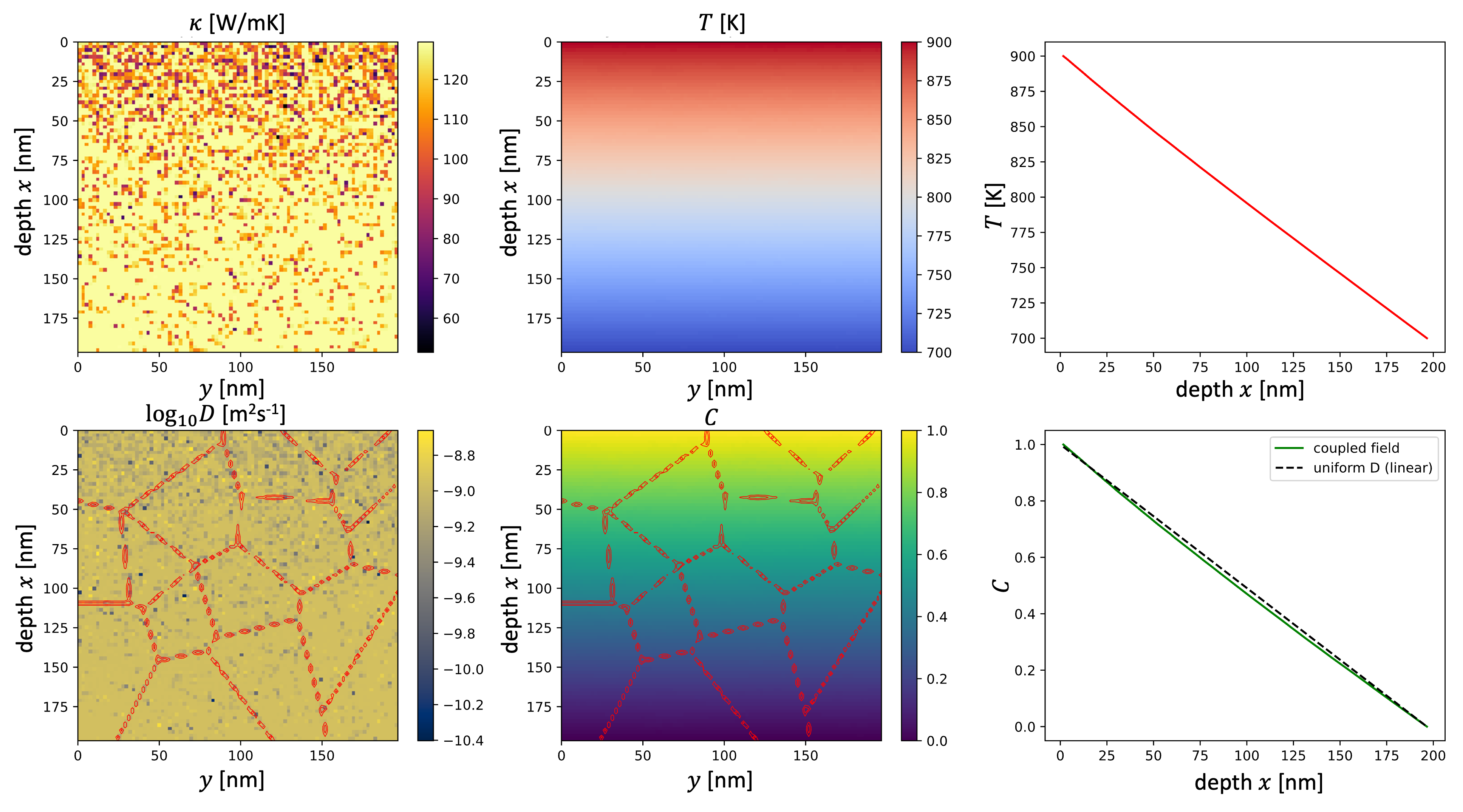}
\caption{Heat and hydrogen transport on the fields obtained by direct
block-wise GNN evaluation of the relaxed $200$\,nm atomistic volume. Left:
$\kappa$ and $D$ fields (grain-boundary network as red contours; dark cells
are individual helium bubbles). Middle: temperature and normalised hydrogen
concentration. Right: depth profiles and effective coefficients.}
\label{fig:field}
\end{figure}

The in-domain block predictions are then distilled into a smooth
\emph{constitutive closure}: $\log_{10}D_0$, $E_a$ and $\kappa_{ph}$ are
fitted (linearly, linearly and quadratically) against the block helium
fraction over the $6513$ grain-interior in-domain blocks (red lines in
Fig.~\ref{fig:closure}), with separate multiplicative grain-boundary
factors obtained from paired polycrystalline/single-crystal teacher
volumes: a continuum cell carries a grain-boundary fraction
$g_\mathrm{b}\in[0,1]$ ($g_\mathrm{b}=0$ for a grain-interior cell,
$g_\mathrm{b}=1$ for a cell lying entirely on a boundary), which at
$g_\mathrm{b}=1$ multiplies $D_0$ by $0.60$ and the phonon conductivity by
$0.54$, interpolating linearly in between. The closure plays two roles. First, it substitutes for the
out-of-domain bubble-interior blocks ($11.5\%$ of blocks): these are
assigned the fit evaluated at the domain limit $f=0.13$ (the fit is never
extrapolated beyond its fitting range), reduced by the Maxwell
effective-medium obstruction factor in the measured per-block void
fraction (treating the bubble interior as an impermeable obstacle for both
channels). Second, it is the constitutive law carried to the component
scale in Sec.~\ref{sec:component}. With the bubble-interior blocks so
treated, the resulting fields are healthy everywhere (positive $D$ and
$\kappa$, no super-pristine values), resolve individual bubbles and the
grain-boundary network (Fig.~\ref{fig:field}), and reproduce the
He--$\kappa$ anticorrelation ($r=-1.0$) of the training data. Homogenising
the volume gives effective coefficients
$\kappa_\mathrm{eff}/\kappa_\mathrm{pristine}=0.94$ (the reduction is modest
because helium is concentrated in a thin near-surface layer) and
$D_\mathrm{eff}/D_\mathrm{pristine}=0.46$, the reduction coming from the
grain-boundary network and from the bubbles acting as discrete impermeable
obstacles.

\subsection{Component-scale extension: coupled monoblock analysis}
\label{sec:component}
\begin{figure}[!tbp]
\centering
\includegraphics[width=0.5\linewidth]{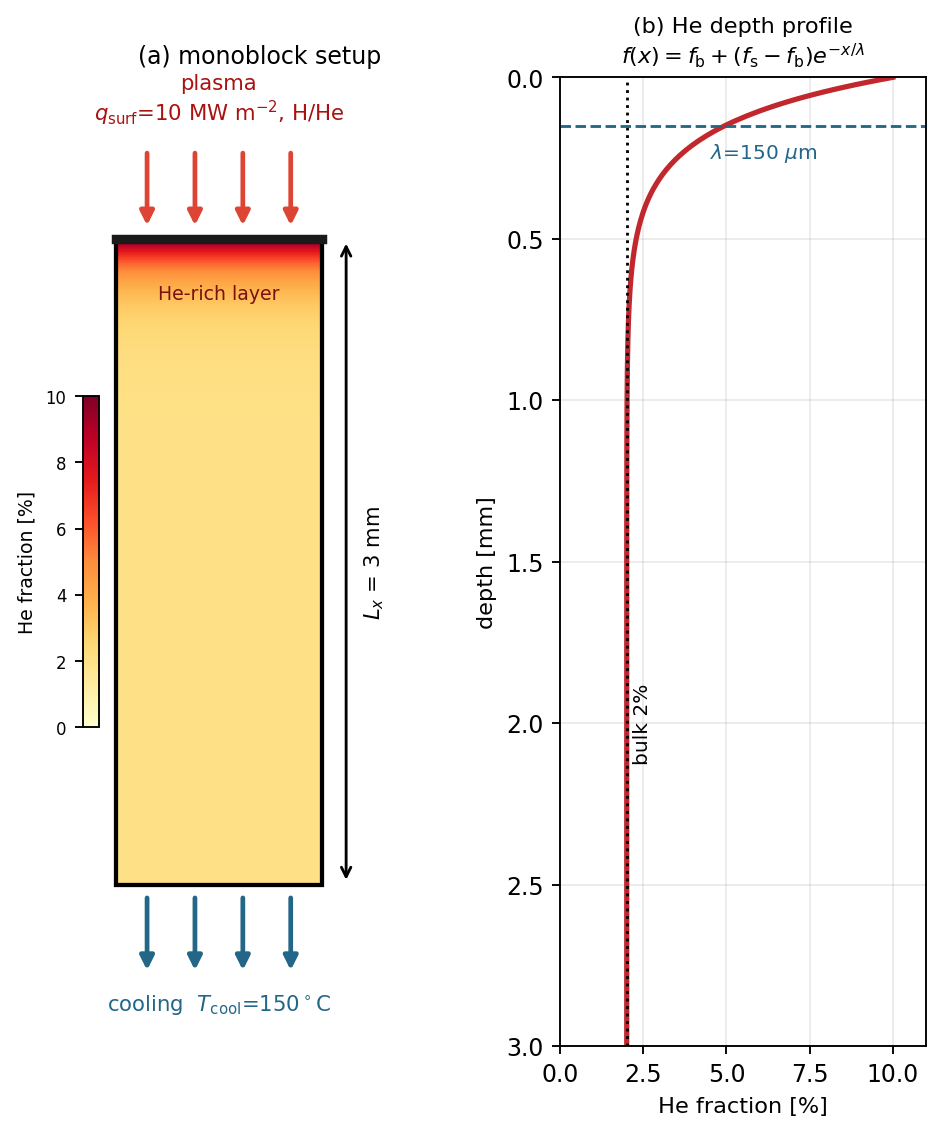}
\caption{Setup of the component-scale monoblock analysis. (a) A 3\,mm
helium-bubble polycrystalline tungsten monoblock: the plasma-facing surface
($x=0$) receives a heat flux $q_\mathrm{surf}=10$\,MW\,m$^{-2}$ and
hydrogen/helium implantation, the back side is cooled at
$T_\mathrm{cool}=150\,^\circ$C, and the shading shows the local helium
fraction (the polycrystalline grain structure, of grain size $\ll$\,mm, is
not drawn and enters through the grain-boundary factor of the closure).
(b) Helium depth profile
$f(x)=f_\mathrm{b}+(f_\mathrm{s}-f_\mathrm{b})e^{-x/\lambda}$ with
implantation range $\lambda=150\,\mu$m and bulk transmutation floor
$f_\mathrm{b}=2\%$.}
\label{fig:setup}
\end{figure}

\begin{figure}[!tbp]
\centering
\includegraphics[width=0.85\linewidth]{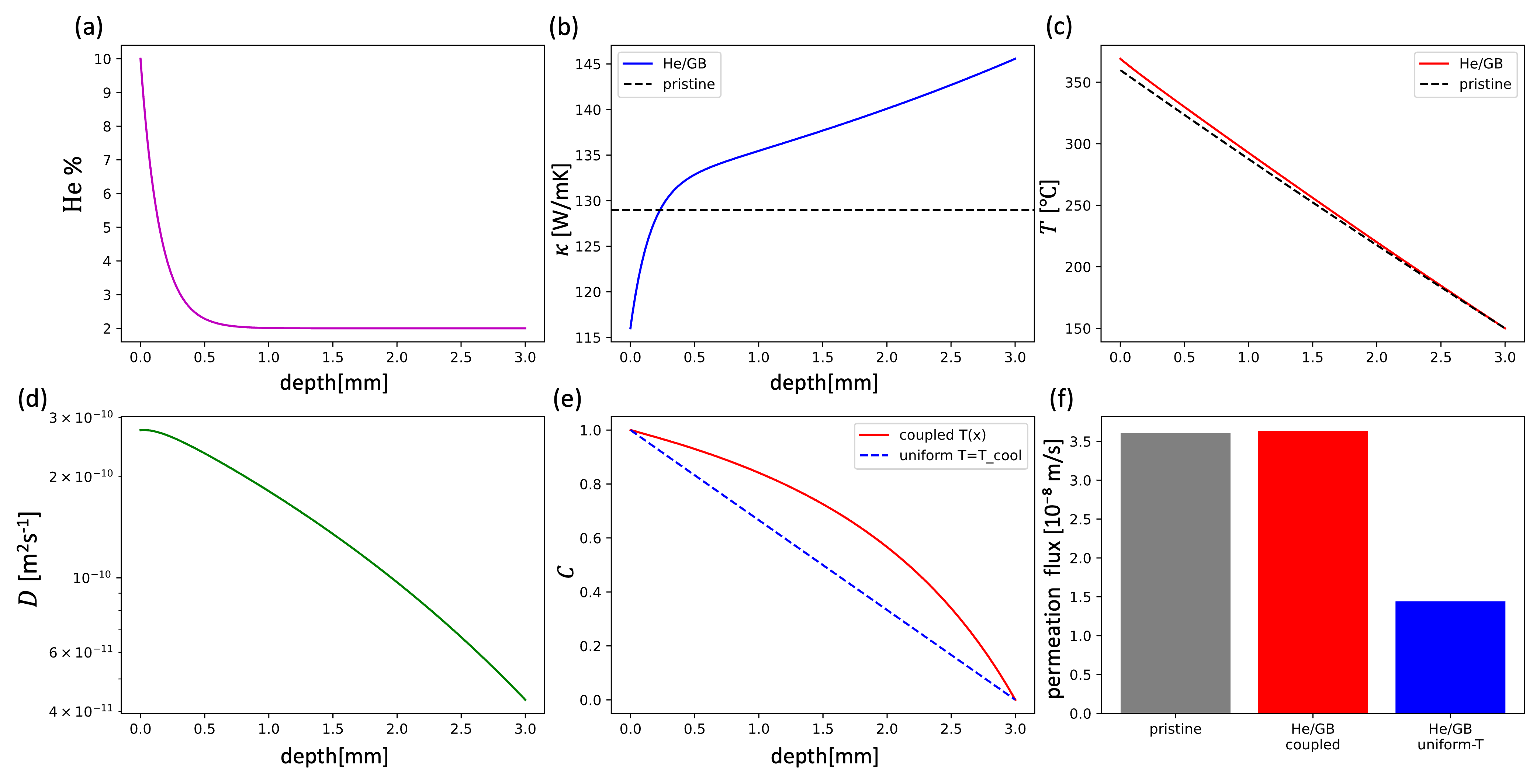}
\caption{Coupled thermal--hydrogen analysis of a 3\,mm tungsten monoblock.
The helium-degraded conductivity raises the temperature (c), which through
$D(T)$ reshapes the hydrogen concentration (e) and permeation (f) relative
to a uniform-temperature assumption.}
\label{fig:coupled}
\end{figure}

The component-scale setting is summarised in Fig.~\ref{fig:setup}: a 3\,mm
tungsten monoblock under a plasma-facing heat flux
$q_\mathrm{surf}=10$\,MW\,m$^{-2}$ with back-side cooling at
$T_\mathrm{cool}=150\,^\circ$C, and a helium fraction that decays from
$\sim$10\% at the implantation surface to a $\sim$2\% bulk transmutation
level over an implantation range $\lambda=150\,\mu$m. At this scale no
atomistic structure exists; the closure of
Sec.~\ref{sec:nanofield} therefore acts as the constitutive law, mapping
the continuum helium profile of Fig.~\ref{fig:setup}(b) and the
grain-boundary fraction to local $\kappa(x,T)$ and $D(x,T)$ in every finite
element. The conductivity degradation raises the surface temperature by
$\sim\!10$\,K.

The central result enabled by having both tensors from the
same structure is the one-way thermal-to-hydrogen coupling: evaluating
$D(T(x))$ on the self-consistent temperature field, rather than assuming a
single representative temperature, changes the predicted permeation flux by a
factor of $\sim\!2.5$ (Fig.~\ref{fig:coupled}), because $D$ depends
exponentially on the $\kappa$-controlled temperature. The \emph{direct}
effect of the helium layer on the steady-state hydrogen inventory is, by
contrast, marginal (ratio $\approx\!1.0$ under both fixed-concentration and
fixed-flux boundary conditions): at divertor-relevant temperature the
block-level reduction of $D$ by moderate helium fractions is modest
(Fig.~\ref{fig:closure}), and bubble trapping manifests instead in the
increased migration barrier of dense-bubble structures at low temperature
(Fig.~\ref{fig:valid}). Helium therefore acts on component-scale hydrogen
transport chiefly through the thermal channel, the
$\kappa\!\to\!T\!\to\!D(T)$ coupling that only a model supplying both
tensors from the same structure can capture.

\section{Discussion}
\label{sec:discussion}
The surrogate learns the two structural mechanisms that dominate divertor
transport: helium bubbles and grain boundaries reduce the effective hydrogen
diffusivity by trapping (a rising activation energy and falling
$D_0$), and the same features degrade conductivity: helium principally
through porosity of the electron-dominated channel and grain boundaries
through phonon scattering. Because the geometric pooling layer makes the
tensor outputs rotation covariant, the model can be queried on arbitrarily
oriented sub-blocks of a large specimen, which is what enables the
microstructure-resolved continuum fields of Sec.~\ref{sec:application}.

Several limitations bound the present absolute accuracy. (i) The bulk labels
are generated with an embedded-atom potential; its description of
H--vacancy binding is weak, which is why isolated vacancies underpredict
trapping relative to grain boundaries and bubbles. The first-principles
anchor itself does not share this deficiency: it reproduces the
mono-vacancy binding energy ($1.15$\,eV, within the 1.0--1.4\,eV spread
of zero-point-corrected sequential DFT binding energies
\cite{HeinolaVH2010}), and the class-resolved anchor calibration of
Sec.~\ref{sec:anchor} therefore corrects the EAM bias at its source. A
trap-augmented refit of the anchor potential on the newly collected
59-structure DFT set was evaluated: it matches the static DFT
benchmarks more closely (a hydrogen migration barrier of $0.20$\,eV),
but its dynamic diffusivities exceed the experimental gas-loading
scale by a factor of 2--6, and the experimentally consistent
production anchor is retained (Sec.~\ref{sec:mlp}). (ii) The Green--Kubo
phonon conductivity from short simulations of small cells is noisy and
contributes only a few percent of the tungsten total; we therefore rely on
the literature-calibrated electronic model for the absolute conductivity and
treat helium bubbles as porosity, which is appropriate for void-type bubbles
but distinct from the dissolved-helium/displacement-damage regime of the
implantation experiments. (iii) The homogenised effective coefficients depend
on the assumed microstructure geometry; the framework itself is geometry
agnostic and consumes whatever field the microstructure generator provides.
The coupling in Sec.~\ref{sec:application} is one-way (heat to hydrogen),
justified because the hydrogen concentration has negligible feedback on
$\kappa$; a two-way outer iteration is a straightforward extension.

\section{Conclusions}
\label{sec:conclusions}
We have built a rotation-equivariant graph neural network that maps a
tungsten atomic configuration containing helium bubbles and grain boundaries
directly to the temperature-dependent $3\times3$ tensors of the hydrogen
diffusion coefficient and thermal conductivity, and returns them in
milliseconds. Trained on 635 microstructures with an embedded-atom teacher,
electronically calibrated against irradiation-degradation data and anchored
to a first-principles machine-learning potential, the surrogate reproduces
literature hydrogen migration barriers (median $E_a=0.21$\,eV) and, following
a dedicated trapping MD campaign, the increase of the barrier and the
$\sim\!2\times$ reduction of effective diffusivity at grain boundaries and
helium bubbles. Coupled finite-element thermal--hydrogen analyses driven by
the surrogate fields show that the conductivity degradation and the resulting
temperature field change predicted hydrogen permeation by a factor of $2.5$,
while the direct effect of the helium layer on the steady-state inventory is
marginal: helium acts on component-scale hydrogen transport chiefly through
the thermal channel, a quantity inaccessible to scalar handbook coefficients.
The class-resolved first-principles anchor correction is already folded into
the released model, and the anchor reproduces the mono-vacancy hydrogen
binding energy ($1.15$\,eV). A trap-augmented refit of the anchor
potential on the collected 59-structure DFT set was evaluated and
deliberately not adopted: it reproduces the static DFT benchmarks but
overshoots the experimental hydrogen diffusivity scale, so the
experimentally consistent production anchor is retained; a dynamically
validated refit that further tightens the vacancy-trapping channel
remains future work.

\section*{CRediT authorship contribution statement}
\textbf{S.~Saito:} Conceptualization, Methodology, Software, Investigation,
Formal analysis, Visualization, Writing -- original draft.
\textbf{M.I.~Kobayashi:} Conceptualization, Resources, Supervision,
Writing -- review \& editing.
\textbf{T.~Kasahara:} Validation.

\section*{Declaration of competing interest}
The authors declare that they have no known competing financial interests or
personal relationships that could have appeared to influence the work
reported in this paper.

\section*{Data availability}
The trained predictor, the inference and demonstration code, and the teacher
labels used for training will be made openly available on GitHub
(\url{https://github.com/saitos-lab/whe-tensor-gnn}) and archived as a
citable snapshot on Zenodo upon publication of this article; until then they
are available from the corresponding author on reasonable request. Scripts
for generating new teacher data (EAM molecular dynamics) and for the FLARE
anchor calibration are available from the corresponding author.

\section*{Declaration of generative AI and AI-assisted technologies in the
writing process}
During the preparation of this work the authors used Claude Code (Anthropic)
in order to assist with English-language editing and the structural
organisation of the manuscript. After using this tool, the authors reviewed
and edited the content as needed and take full responsibility for the
content of the published article.

\section*{Acknowledgements}
This study is based on the methodology developed under the NIFS
Collaborative Research Program (NIFS25KSPT009, NIFS24KIPT013,
NIFS25KIST066, and NIFS22KIGS002). S.~Saito, M.I.~Kobayashi and T.~Kasahara are supported by JST (Moonshot
R\&D Program), Japan, Grant Number JPMJMS24A3. The computations were
performed using the Plasma Simulator of NIFS (Toki, Gifu, Japan).

\bibliographystyle{elsarticle-num}

\end{document}